\documentclass[a4paper]{jpconf}

\usepackage{graphicx}
\usepackage{citesort}
\begin{document}
\nocite{*}
\title{Measurement of $b$-baryons with the CDF II detector}

\author{Joachim Heuser for the CDF Collaboration}

\address{Institut f\"ur Experimentelle Kernphysik, University of Karlsruhe, Wolfgang-Gaede-Str. 1, 76131 Karlsruhe, Germany}

\ead{joachim.heuser@ekp.uka.de}

\begin{abstract}
We report the observation of new bottom baryon states. The most recent
result is the observation of the baryon $\Xi_b^{-}$ through the decay
 $\Xi_b^{-} \to J/\psi \Xi^{-}$. The significance of the signal corresponds
to 7.7$\sigma$ and the $\Xi_b^{-}$ mass is measured to be
$5792.9 \pm 2.5\mathrm{(stat.}) \pm 1.7\mathrm {(syst.)\; MeV/c^2}$. In addition
we observe four resonances in the $\Lambda_b^0 \pi^{\pm}$ spectra, consistent
with the bottom baryons $\Sigma_b^{\mathrm{(*)\pm}}$. All observations are in
agreement with theoretical expectations.
\end{abstract}

\section{Introduction}
The quark model has been very successful in describing the spectroscopy
of hadrons, both for light hadrons as well as for hadrons with heavy quarks.
The spectroscopy of heavy baryons (or mesons) provides an interesting
laboratory for understanding the theory of strong interactions,
Quantum Chromodynamics (QCD), in a regime where
perturbation calculations cannot be applied. In effective models of
the heavy hadron systems, like heavy quark effective theory
(HQET) \cite{Manohar:1993qn}, the degrees of freedoms of the
heavy quark are considered decoupled
from those of the light quarks, so that a heavy baryon system
can be modeled in a similar way as the helium atom is modeled.\\
Experimental results in the $b$-baryon sector have so far been
limited to one single state, the $\Lambda_b^0$ with quark content (udb).
In
these proceedings we present the observation and the mass measurement of
further $b$-baryon states: the $\Xi_b^{-}$ state \cite{cdf_xib_observation:2007un}
and the $\Sigma_b^{\mathrm{(*)\pm}}$ states \cite{cdf_sigmab_observation:2007rw}.\\

\section{Observation of the bottom baryon {$\Xi_b^{-}$} }
\begin{figure}
\includegraphics[width=22pc]{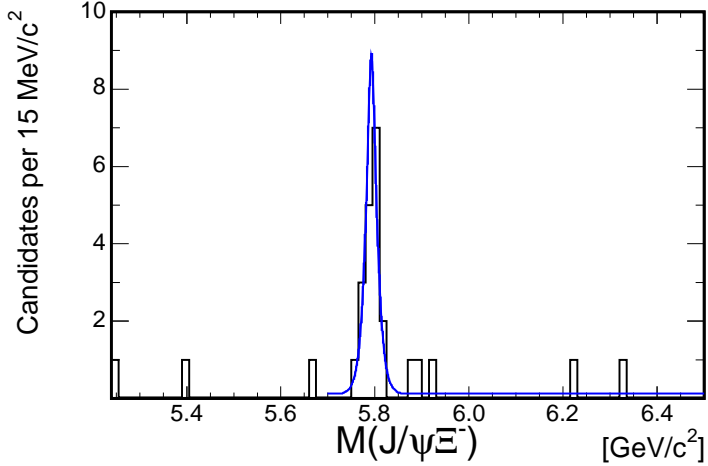}\hspace{2pc}%
\begin{minipage}[b]{12pc}\caption{\label{Xi_b_spectrum}The mass distribution
of $\Xi_b^{-}$ candidates after cut optimization. Also shown is the projection of the
used fit function, yielding $17.5 \pm 4.3$ $\Xi_b^{-}$ candidates.}
\end{minipage}
\end{figure}

The baryon with quark content (dsb) and spin $S=\frac{1}{2}$
is labelled {$\Xi_b^{-}$} in the baryon naming scheme.
Using $1.9 \; {\mathrm{fb^{-1}}}$ of data collected
with the CDF II detector, {$\Xi_b^{-}$} candidates are
reconstructed in the decay chain $\Xi_b^{-} \to J/\psi \; \Xi^-$,
where $J/\psi \to \mu^{+}\mu^{-}$, $\Xi^- \to \Lambda \pi^-$, and
$\Lambda \to p \pi^-$.\\
An important feature of the analysis is that the intermediate
$\Xi^-$ baryon can be tracked by precision measurements
in the silicon layers of the CDF II detector, since 
the $\Xi^-$ is a charged and long-lived particle. This
significantly improves the
secondary vertex resolution and strongly helps to suppress background of
random $\Lambda \pi^-$ combinations.\\
It is expected that the mass splitting between the $b$-baryons
$\Lambda_b$ and $\Xi_b$ is similar to that between the $c$-baryons
$\Lambda_c$ and $\Xi_c$, leading to an expected value of $\sim 5.8\; \mathrm{GeV/c^2}$
for the $\Xi_b$ mass \cite{Jenkins:1996de,Ebert:2005xj}.
Furthermore the decay properties should be dominated by the weak
transition of the $b$-quark, so that the decay of the $\Xi_b$ should
show similarities to those of other weakly decaying $b$-hadrons.\\
The last fact is exploited to choose an unbiased selection procedure of the
$\Xi_b^{-}$-candidates. 
A sample of $\sim 30,000$ $B^+ \to J/\psi K^+$ decays, which are
kinematically similar to the desired $\Xi_b^{-} \to J/\psi \Xi^-$ decays,
is used to optimize the selection.
The
result is shown in Fig. \ref{Xi_b_spectrum}. A clear signal is
visible and its mass is measured to be
$5792.9 \pm 2.5\mathrm{(stat.}) \pm 1.7\mathrm {(syst.)\; MeV/c^2}$.
This is in good agreement with a recent measurement from
D0 \cite{d0_xib_observation:2007un} and with theory predictions.
The probability to observe a background fluctuation of this size is evaluated to be
$6.6 \times 10^{-15}$, corresponding to a signal significance of $7.7\sigma$.

\section{Observation of the bottom baryon states $\Sigma_b^{\pm}$ and $\Sigma_b^{\pm*}$ }
The charged {$\Sigma_b$} baryon states have quark content (uub) and
(ddb). In HQET, the light diquark system, treated separately from the $b$-quark,
has isospin $I=1$ and spin $j=1$. Together with the $b$-quark the light quarks form
the isospin triplet $\Sigma_b^{+}$, $\Sigma_b^{0}$, $\Sigma_b^{-}$
(the corresponding isospin singlet baryon state is the
$\Lambda_b^{0}$). The spin $j=1$ of the diquark system can couple
with that of the $b$-quark to either $J=\frac{1}{2}$ or $J=\frac{3}{2}$.
The triplet states with $J=\frac{1}{2}$ form the ground state $\Sigma_b$ baryons,
while the states with $J=\frac{3}{2}$ are labelled $\Sigma_b^{*}$.
The range of theoretical predictions for the expected masses is shown
in Tab. \ref{Sigma_b_predictions}.

\begin{table}[h]
\begin{minipage}{18pc}
\caption{\label{Sigma_b_predictions}Mass and width predictions for the $\Sigma_b^{\pm(*)}$. See
\cite{cdf_sigmab_observation:2007rw} for an extensive list of references.}
\begin{center}
\lineup
\begin{tabular}{lc}
\br
Quantity &  ($\mathrm{MeV/c^2}$) \\
\mr
$m(\Sigma_b)-m(\Lambda_b^0)$     & 180 -- 210  \\
$m(\Sigma_b^{*})-m(\Sigma_b)$    & 10 -- 40    \\
$m(\Sigma_b^{-})-m(\Sigma_b^{+})$ & 5 -- 7      \\
\mr
$\Gamma(\Sigma_b)$, $\Gamma(\Sigma_b^*)$ & $\sim 8$, $\sim 15$ \\
\br
\end{tabular}
\end{center}
\end{minipage}\hspace{2pc}%
\begin{minipage}{18pc}
\caption{\label{Sigma_b_measurement}Measured masses for the $\Sigma_b^{\pm(*)}$ states,
calculated from the Q values with $m(\Lambda_b^0)$ from \cite{Acosta:2005mq}. }
\begin{center}
\begin{tabular}{ll}
\br
State &  Mass ($\mathrm{MeV/c^2}$) \\
\mr
$\Sigma_b^{+}$  &  $5807.8^{+2.0}_{-2.2}\mathrm{(stat.})\pm{1.7}\mathrm{(syst.})$ \\
$\Sigma_b^{-}$  &  $5815.2^{+1.0}_{-1.0}\mathrm{(stat.})\pm{1.7}\mathrm{(syst.})$ \\
$\Sigma_b^{*+}$ &  $5829.0^{+1.6}_{-1.8}\mathrm{(stat.})^{+1.7}_{-1.8}\mathrm{(syst.})$ \\
$\Sigma_b^{*-}$ &  $5836.4^{+2.0}_{-2.0}\mathrm{(stat.})^{+1.8}_{-1.7}\mathrm{(syst.})$ \\
\br
\end{tabular}
\end{center}
\end{minipage}
\end{table}

\begin{figure}[h]
\begin{minipage}{18pc}
\includegraphics[width=18pc]{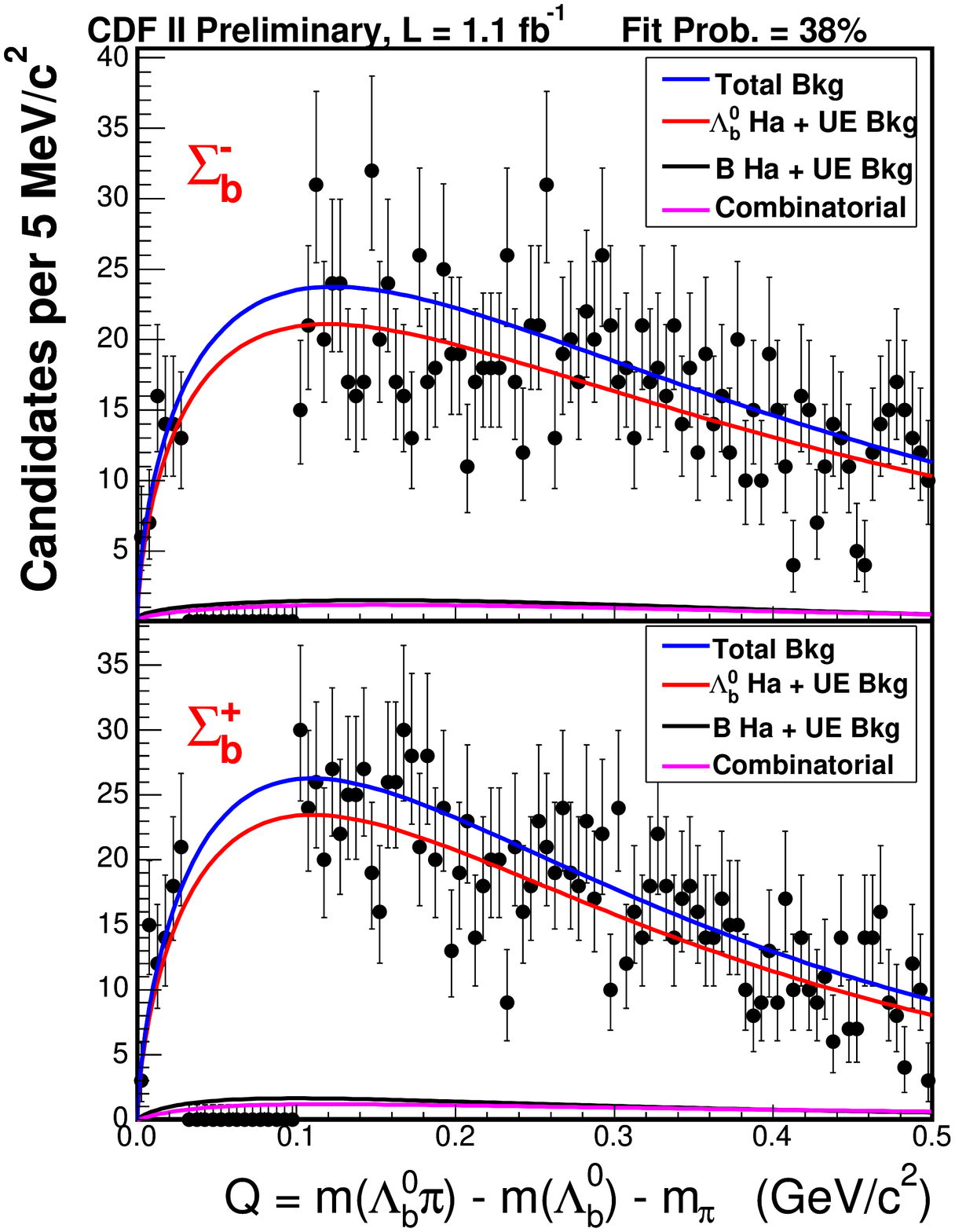}
\caption{\label{sigma_spectrum_blinded}The Q spectra and the background estimation of
the $\Sigma_b^{-(*)}$ (top figure) and $\Sigma_b^{+(*)}$ (bottom figure) candidates
after cut optimization.}
\end{minipage}\hspace{2pc}%
\begin{minipage}{18pc}
\includegraphics[width=18pc]{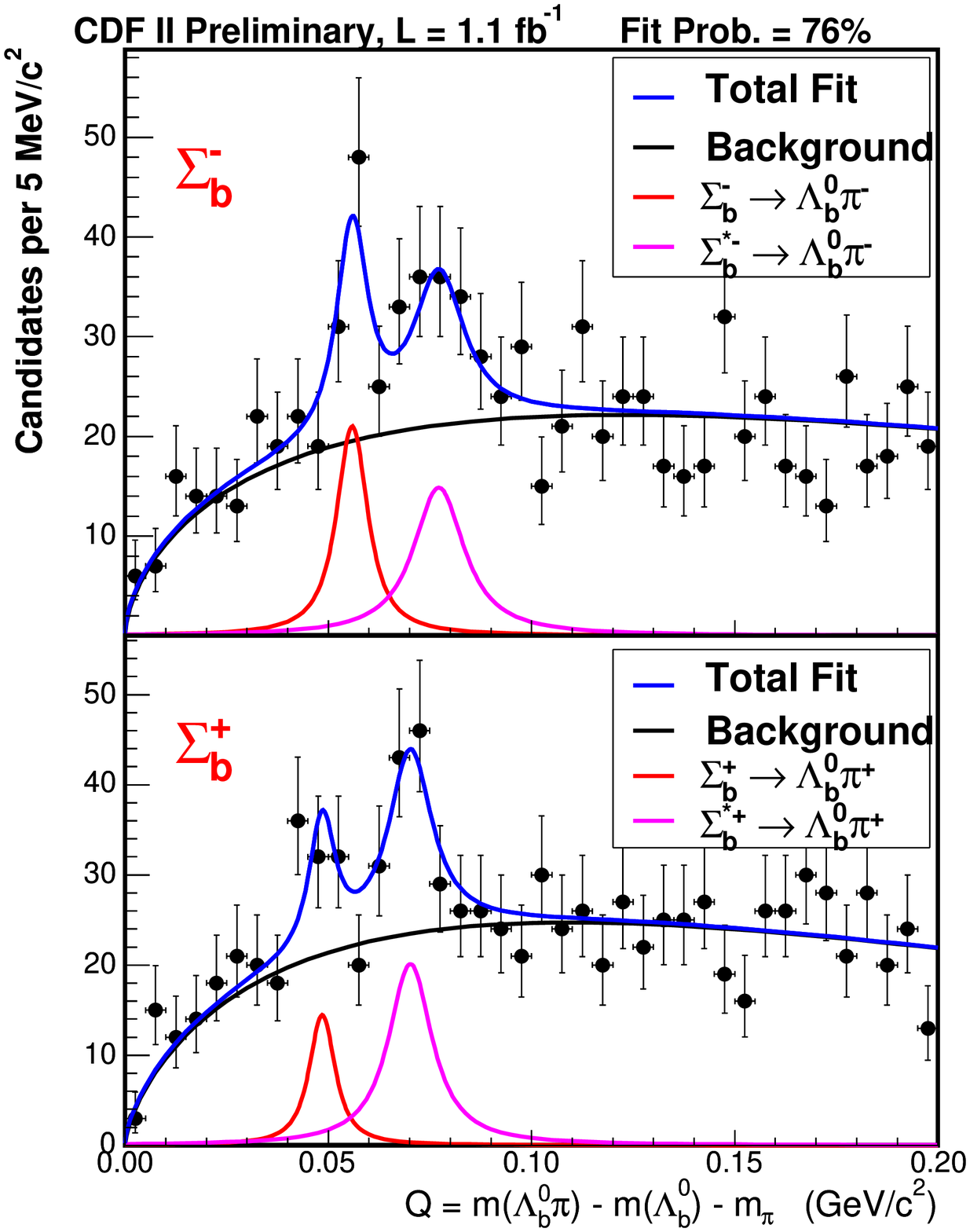}
\caption{\label{sigma_spectrum_fit}The fit to the Q spectra after unblinding the signal
region.}
\end{minipage} 
\end{figure}

The search is based on $1.1 \; {\mathrm{fb^{-1}}}$ of data using the decay mode
$\Sigma_b^{\pm(*)} \to \Lambda_b^0 \pi^{\pm}$, where
$\Lambda_b^0 \to \Lambda_c^+ \pi^-$ and $\Lambda_c^+ \to p K^{-} \pi^+$. 
A sample with $\sim 3200$ $\Lambda_b^0$ baryons is combined with charged pion tracks
to obtain the $\Sigma_b^{\pm(*)}$ candidates. The search is performed in
the variable $Q = m(\Lambda_b^0\pi^\pm) - m(\Lambda_b^0) - m(\pi^\pm)$ to
minimize the contribution of the mass resolution of each
$\Lambda_b^0$ candidate. During the cut optimization process and the
determination of the
background contributions, the signal region, estimated from
theory predictions, is kept blinded (see Fig. \ref{sigma_spectrum_blinded}).\\
After unblinding the spectrum, an excess is observed in the signal region.
The $\Sigma_b^{-(*)}$ and $\Sigma_b^{+(*)}$ spectra are fitted
simultaneously with an unbinned maximum likelihood fit, where
$m(\Sigma_b^{+*})-m(\Sigma_b^+)$ is constrained to be identical to
$m(\Sigma_b^{-*})-m(\Sigma_b^-)$. The projection of the fit result is shown
in Fig. \ref{sigma_spectrum_fit} and the measured $\Sigma_b^{\pm(*)}$ masses
are listed in Tab. \ref{Sigma_b_measurement}.
The null hypothesis (no signal) is excluded by more than five standard
deviations and, except for the $\Sigma_b^+$ signal, each single signal has a
significance exceeding three standard deviations. 


\section{Conclusions}
In summary, the CDF Collaboration has observed both the four lowest-lying charged
$\Sigma_b^{\pm(*)}$ baryons as well as the negatively charged $\Xi_b^{-}$ baryon. All
results are in good agreement with theoretical predictions.

\section*{References}
\bibliography{EPS_baryons_CDF_bib}

\end{document}